\documentclass[
showpacs,
floatfix,
aps,
pra,
amsmath,
twocolumn,
superscriptaddress,
altaffillsymbol
]{revtex4-1}
\usepackage{graphicx}  
\usepackage{dcolumn}   
\usepackage{bm}        
\usepackage{amssymb}   
\usepackage{verbatim}

\usepackage{amsmath}
\usepackage{amsfonts}
\usepackage{booktabs}
\usepackage{multirow}

\def\lsim{\mathrel{\rlap{\lower4pt\hbox{\hskip1pt$\sim$}}
    \raise1pt\hbox{$<$}}}                
\def\gsim{\mathrel{\rlap{\lower4pt\hbox{\hskip1pt$\sim$}}
    \raise1pt\hbox{$>$}}}                

\begin{document}

\title{Origin and Suppression of $1/f$ Magnetic Flux Noise}


\author{P. Kumar}
\author{S. Sendelbach}
\altaffiliation[Present address: ]{Northrop Grumman Corporation, Linthicum, Maryland 21203, USA}
\author{M. A. Beck}
\affiliation{Department of Physics, University of Wisconsin-Madison, Madison, Wisconsin 53706, USA}
\author{J. W. Freeland}
\affiliation{Advanced Photon Source, Argonne National Laboratory, Argonne, Illinois 60439, USA}
\author{Zhe Wang}
\affiliation{Department of Physics and Astronomy,
University of California, Irvine, California 92617, USA}
\affiliation{State Key Laboratory of Surface Physics and Key Laboratory
for Computational Physical Sciences, Fudan University, Shanghai 200433, China}
\author{Hui Wang}
\affiliation{Department of Physics and Astronomy,
University of California, Irvine, California 92617, USA}
\affiliation{State Key Laboratory of Surface Physics and Key Laboratory
for Computational Physical Sciences, Fudan University, Shanghai 200433, China}
\author{C. C. Yu}
\author{R. Q. Wu}
\affiliation{Department of Physics and Astronomy,
University of California, Irvine, California 92617, USA}
\author{D. P. Pappas}
\affiliation{National Institute of Standards and Technology, Boulder, Colorado 80305, USA}
\author{R. McDermott}
\email[Electronic address: ]{rfmcdermott@wisc.edu}
\affiliation{Department of Physics, University of Wisconsin-Madison, Madison, Wisconsin 53706, USA}

\date{\today}

\begin{abstract}
 Magnetic flux noise is a dominant source of dephasing and energy relaxation in superconducting qubits. The noise power spectral density varies with frequency as $1/f^\alpha$ with $\alpha \lsim 1$ and spans 13 orders of magnitude. Recent work indicates that the noise is from unpaired magnetic defects on the surfaces of the superconducting devices. Here, we demonstrate that adsorbed molecular O$_2$ is the dominant contributor to magnetism in superconducting thin films. We show that this magnetism can be suppressed by appropriate surface treatment or improvement in the sample vacuum environment. We observe a suppression of static spin susceptibility by more than an order of magnitude and a suppression of $1/f$ magnetic flux noise power spectral density by more than a factor of 5. These advances open the door to realization of superconducting qubits with improved quantum coherence.
\end{abstract}

\maketitle

A quantum computer will allow efficient solutions for certain problems that are intractable on conventional, classical computers, including factoring and quantum simulation. Superconducting quantum bits (``qubits'') based on Josephson junctions are a leading candidate for scalable quantum information processing in the solid state \cite{Devoret04, Clarke08}. Gate and measurement operations have attained a level of fidelity that should enable quantum error correction \cite{Kelly15, Corcoles15}, and there is interest in scaling to larger systems \cite{Fowler12, Gambetta15}. However, qubit performance is limited by dephasing \cite{Martinis03, Ithier05}. The dominant source of dephasing is low-frequency $1/f$ magnetic flux noise \cite{Yoshihara06, Kakuyanagi07, Bialczak07}. Uncontrolled variation of the flux bias of the qubit leads to the accumulation of spurious phase during periods of free evolution, resulting in a rapid decay of qubit coherence. Magnetic flux noise was first identified in the 1980s \cite{Wellstood87, Weissman88}. Efforts to avoid flux noise include operation at a ``sweet spot'' where the device is insensitive to first order to magnetic flux fluctuations \cite{Vion02}, or elimination of superconducting loops that allow the frequency of the qubit to be tuned \textit{in situ} \cite{Paik11}. However, restriction to fixed-frequency qubits results in longer gate times, and static disorder in the junction critical currents makes it difficult to target specific frequencies, leading to frequency clashes in larger multiqubit circuits. In the context of a quantum annealer \cite{Johnson11, Boxio16}, flux noise degrades performance by limiting the number of qubits that can tunnel coherently. For these reasons, there is strong motivation to understand and eliminate the flux noise.

Recent experiments indicate that there is a high density of unpaired surface spins in superconducting integrated circuits \cite{Sendelbach08} and it is believed that fluctuations of these spins give rise to the $1/f$ flux noise \cite{Faoro08,Wang15,LaForest15}. There is experimental evidence that interactions between the surface spins are significant \cite{Sendelbach09}. To date, however, there has been no experimental data pointing toward the microscopic nature of the surface magnetic defects, although there has been speculation that the defects are due to localized states at the disordered metal-insulator interface \cite{Choi09} or to surface adsorbates \cite{Lee14}, in particular molecular O$_2$ \cite{Wang15}.

\begin{figure}[t!]
\includegraphics[width=.47\textwidth]{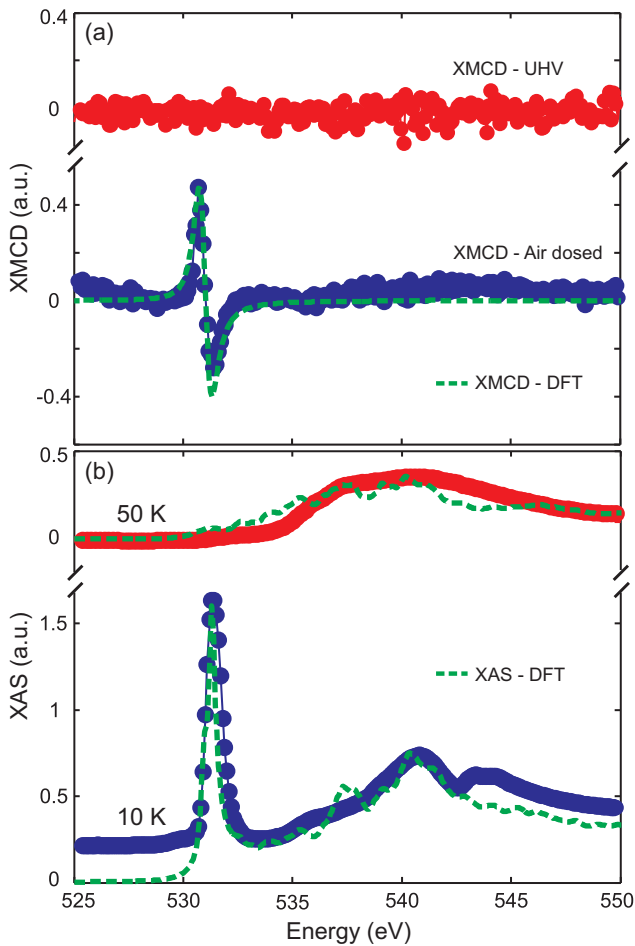}
\vspace*{-0.0in} \caption{(a) Xray magnetic circular dichroism (XMCD) at the oxygen K-edge for a native Al film and an Al film exposed to air. The native film (top) shows no XMCD signal, while the air-exposed film (bottom) shows a clear XMCD signal at 531~eV. (Traces are offset for clarity). (b) Oxygen K-edge Xray absorption spectroscopy (XAS) of an Al thin film cooled in the presence of $5\times 10^{-8}$~Torr O$_2$. Beginning around 45 K we observe a sharp peak at 531~eV and a broad spectral feature from 540-545~eV which we ascribe to adsorbed molecular O$_2$. (Traces are offset for clarity). Dashed lines are from DFT simulations for Al$_2$O$_3$ (XAS at 50 K) and for O$_2$/Al$_2$O$_3$ (XMCD and XAS at 10 K); see Supplement for details.}
\label{fig:fig1}\end{figure}

Here we describe a series of Xray Absorption Spectroscopy (XAS) and Xray Magnetic Circular Dichroism (XMCD) experiments that point to adsorbed molecular O$_2$ as the dominant source of surface magnetism in superconducting thin films. We show that improvement in the vacuum environment of the superconducting sample, in conjunction with appropriate surface passivation to inhibit adsorption of residual magnetic species, can dramatically reduce the surface density of spins in superconducting thin films. We present data on surface spin susceptibility and magnetic flux noise of devices before and after various surface treatments and demonstrate a significant suppression of magnetic activity and flux noise power.  These results represent a major step forward in a decades-old problem in condensed matter and device physics and open the door to realization of improved frequency-tunable qubits with extended dephasing times.

Using the Advanced Photon Source (APS) at Argonne National Laboratory, we have performed XAS and XMCD experiments on aluminum and niobium thin film samples. In XMCD, one monitors the absorption of a spin-polarized sample at specific Xray absorption edges; the Xray energy provides elemental specificity, while the Xray helicity provides access to the orbital magnetism of the sample. Devices were cooled in a $^4$He flow cryostat to 10 K, and XMCD experiments were performed in magnetic fields up to 5 T. Initially we examined sputtered Al and Nb films cooled in ultrahigh vacuum (UHV; $P\lsim 10^{-9}$ Torr); we expect these films to be covered by an amorphous native oxide due to prolonged exposure to atmosphere prior to characterization at the APS. We examined the Al and O K-edges in the Al films and the Nb L-edge and O K-edge in the Nb films and observed no XMCD signal at any of these energies [Fig. \ref{fig:fig1}(a), upper trace]. However, when we intentionally degraded the vacuum of the sample cryostat by bleeding in air or dry O$_2$ gas at a pressure of order $10^{-6}$ Torr for several minutes, we observed a clear XMCD signal at the O K-edge [Fig. \ref{fig:fig1}(a), lower trace]. Density functional theory (DFT) modeling allows us to assign the measured XMCD signal to molecular O$_2$ [dashed line in Fig. \ref{fig:fig1}(a)].  
In a separate series of experiments, we dosed the metal thin film continuously as we cooled down from room temperature in an O$_2$ partial pressure of $5\times10^{-8}$ Torr; the experimental data and corresponding DFT calculations are shown in Fig. \ref{fig:fig1}(b). We observe a strong modification of the O K-edge XAS signal starting at a temperature around 45 K, indicating the onset of significant adsorption. By comparing the spectral weight of the broad feature from 535-550 eV in the high-temperature spectra to that of the narrow peak at 531 eV in the low-temperature spectra, we can roughly quantify the amount of adsorbed oxygen relative to that bound in the native oxide of the metal. We conclude that the films are covered by 1-2 monolayers of adsorbed O$_2$. The best agreement between DFT and the measured XMCD and XAS signals occurs when the O$_2$ bond is tilted with respect to the beam direction. This 
is consistent with prior DFT calculations of O$_2$ adsorbed on Al$_2$O$_3$ (0001), which indicate that the molecular bond axis is tilted at 55$^\circ$ from the surface normal \cite{Wang15}.

The XMCD results suggest that the dominant magnetism in Al and Nb thin films of the type used to make qubit circuits is due not to a high density of intrinsic defects, but rather to adsorbed molecular O$_2$. Due to the bonding of the O$_2$ molecule, the outermost electrons form a spin 1 triplet state \cite{Wang15}. O$_2$ is paramagnetic at high temperature; at low temperature, solid molecular O$_2$ displays a complex phase diagram with multiple competing magnetic orders \cite{Freiman04}. In typical superconducting qubit experiments, devices are cooled to millikelvin temperatures in vacuum cryostats that achieve pressure of order 10$^{-6}$~Torr prior to cooldown; this pressure corresponds to an adsorption rate of roughly 1 ML/s, assuming a unit sticking coefficient. Even when the cryostat is cold, there will be a continual flux of molecules from hot regions of the cryostat to cold regions where the sample is housed. Thus, an accumulation of magnetic O$_2$ on the surface of these devices is inevitable.

\begin{figure}[t!]
\includegraphics[width=.47\textwidth]{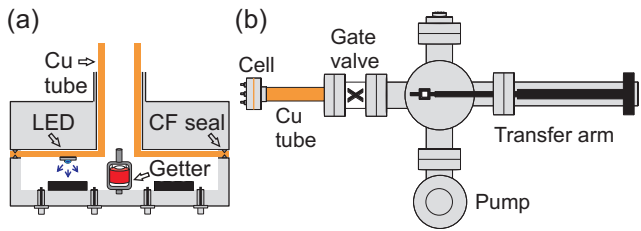}
\vspace*{-0.0in} \caption{(a) Schematic of hermetic grade 5 titanium enclosure for susceptibility and flux noise measurements. The enclosure incorporates weld-in SMA feedthroughs and a single 2.75'' conflat gasket. (b) Schematic of the sample prep chamber. The chamber incorporates a turbo pump, an ion pump, and a transfer arm used to install the NEG in the sample chamber following activation.}
\label{fig:fig2}\end{figure}

This realization motivates an attempt at noise reduction by improving the vacuum environment of the superconducting devices. To this end, we have designed hermetic sample enclosures based on grade 5 titanium alloy (Ti-6Al-4V); see Fig. \ref{fig:fig2}. This alloy has excellent UHV properties due to its low outgassing and its hardness, allowing realization of all-metal conflat seals. Moreover, the material is compatible with high-bandwidth weld-in hermetic SMA connectors. Finally, grade 5 titanium superconducts at around 4.5 K, providing a magnetic shield for sensitive superconducting devices.

In Fig. \ref{fig:fig2}a we show the details of the sample enclosure, and in Fig. \ref{fig:fig2}b we show a schematic of the sample prep chamber. The sample box is pumped through a copper pinch tube with a turbomolecular pump and an ion pump. During evacuation, the sample enclosure and chamber are baked to 120$^\circ$C. Following vacuum bake, the sample cell is cooled to room temperature and the cell is hermetically sealed using a commercial pinch tool. In some cases, the sample cell was backfilled with NH$_3$ gas prior to pinchoff. In other cases, the sample was irradiated with UV light (365 nm) during evacuation to promote photodesorption of strongly bound magnetic species, and a nonevaporable getter (NEG) pill (SAES Inc.) was activated in a separate chamber and transferred into the sample enclosure under vacuum. The NEG provides continuous pumping in the sample cell following pinchoff. 

\begin{figure}[t!]
\includegraphics[width=.47\textwidth]{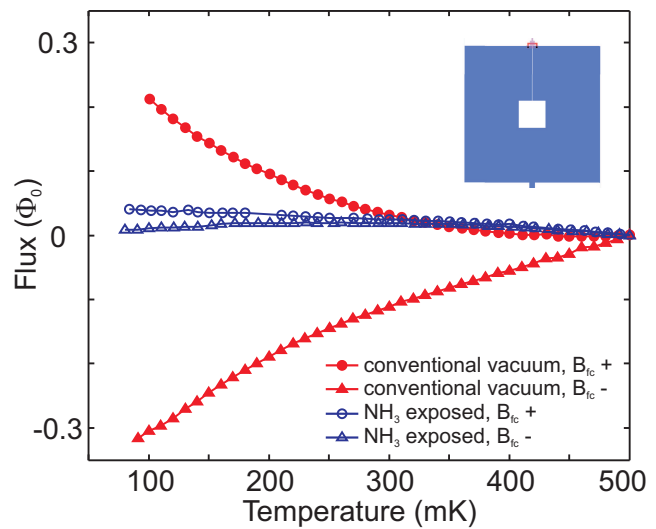}
\vspace*{-0.0in} \caption{Suppression of magnetic susceptibility. Temperature-dependent flux threading a square-washer Nb SQUID (350~pH; see inset) cooled in a conventional vacuum environment (closed red symbols) and cooled following vacuum bake and NH$_3$ passivation (blue open symbols). The upper (lower) branches correspond to cooling fields of +128 $\mu$T (-128 $\mu$T). The magnitude of the flux change is proportional to the density of magnetically active surface spins \cite{Sendelbach08}.}
\label{fig:fig3}\end{figure}

In a first series of experiments, we characterized the surface spin density on washer-style thin-film Nb Superconducting QUantum Interference Devices (SQUIDs) by monitoring the temperature-dependent zero-frequency surface spin susceptibility of field-cooled devices, after the method described in \cite{Sendelbach08}. The device layout is shown in the inset of Fig. \ref{fig:fig3}. Here, we intentionally trap flux vortices in the thin films of the Nb SQUID by cooling through the superconducting transition in the presence of an applied magnetic field; the vortex density is given by $\sigma_v = B_{fc}/\Phi_0$, where $B_{fc}$ is the strength of the cooling field and $\Phi_0 = h/2e$ is the magnetic flux quantum. Any unpaired magnetic defects on the surface of the device develop a thermal polarization in the relatively strong (tens of mT) local magnetic fields in the vortex core. As temperature decreases, the thermal polarization of the defect spins increases, forcing a redistribution of vortex currents; a fraction of these currents circulate around the loop of the SQUID, coupling a flux change to the SQUID loop. The temperature-dependent flux through the SQUID loop thus displays a roughly $1/T$ Curie-like dependence on temperature, and the measured flux change can be used to extract a surface density of unpaired spins. For typical devices, we infer a surface spin density of order 10$^{17}$ m$^{-2}$ \cite{Sendelbach08, Koch07}.

In Fig. \ref{fig:fig3} we compare baseline data to data from a cell that was evacuated and then backfilled with NH$_3$ gas at approximately 100~Torr prior to pinchoff. The temperature-dependent flux is suppressed by roughly an order of magnitude. Nonmagnetic NH$_3$ has a higher free energy of adsorption than O$_2$ (1.5~eV \textit{versus} 0.15~eV according to our DFT calculations on Al$_2$O$_3$), and hence occupies available surface sites that would otherwise be taken up by magnetic O$_2$, resulting in a suppression of the surface density of adsorbed spins; a related approach to suppressing magnetic adsorbates was suggested in \cite{Lee14}.

\begin{figure}[t!]
\includegraphics[width=.47\textwidth]{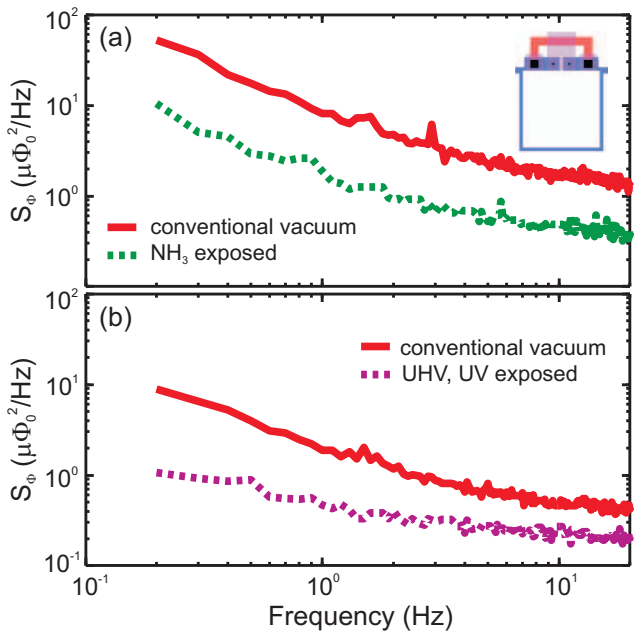}
\caption{(a) Flux noise spectra of SQUID device SiN$_x$-4 before (upper trace) and after (lower trace) vacuum bakeout and NH$_3$ passivation. Inset shows device layout. (b) Flux noise spectra of SQUID device SiN$_x$-6 before (upper trace) and after (lower trace) vacuum bakeout and UV illumination. 
}
\label{fig:fig4}
\end{figure}

Both susceptibility and magnetization noise scale linearly with spin density, and reduction in the density of surface spins should yield a reduction in flux noise power. In a final series of experiments, we have examined the flux noise of Al-based SQUIDs subjected to various surface treatments; the results are presented in Fig. \ref{fig:fig4} and Table I. In these experiments, the Al-based first-stage SQUID is biased with a voltage, and the fluctuating current through the device is measured with a second Nb-based SQUID \cite{Wellstood87}; measurements are performed in an adiabatic demagnetization refrigerator (ADR) at a temperature of 100 mK. By varying the flux bias point of the first-stage device, we can verify that the dominant noise contribution is indeed flux-like and not, e.g., due to critical current fluctuations. We have characterized devices where the SQUID loop is encapsulated either in SiN$_x$ or SiO$_x$ grown by plasma enhanced chemical vapor deposition (PECVD). 
The SQUIDs described here were designed with a relatively high loop aspect ratio (ratio of loop width to trace width) of 25, as this geometry enhances the coupling of surface spin fluctuations to the device \cite{Bialczak07, LaForest15} (see Supplement). We fit the measured noise spectra to the form $A/f^\alpha + B$, and we compare the $1/f$ noise power $A$ and noise exponent $\alpha$ measured on identical devices before and after surface treatment. In all, we have examined before/after spectra of 10 devices.

\begin{table}[t]
\centering
\begin{tabular}{cccccc}
 & & \multicolumn{2}{c}{Pre-treatment} & \multicolumn{2}{c}{Post-treatment} \\
\multirow{2}{*}{Device} &  \multirow{2}{*}{\hspace{6pt} Treatment \hspace{6pt}} & \hspace{2pt} $S_\Phi$(1 Hz) \hspace{2pt} & \multirow{2}{*}{$\alpha$} & \hspace{4pt} $S_\Phi$(1 Hz) \hspace{2pt}& \multirow{2}{*}{$\alpha$} \\
 &  & ($\mu\Phi_0^2$/Hz) & & ($\mu\Phi_0^2$/Hz) & \\
\hline
\hline
SiN$_x$-1 & UHV & 2.0 & 1.0 & 1.4 & 1.1 \\
\hline
SiN$_x$-2 & NH$_3$ & 4.4 & 0.7 & 2.4 & 0.7 \\
\hline
SiN$_x$-3 & UHV,UV & 2.8 & 1.0 & 1.3 & 0.9 \\
\hline
\multirow{2}{*}{SiN$_x$-4} & NH$_3$ &  \multirow{2}{*}{8.2} & \multirow{2}{*}{1.2} & 1.6 & 1.1 \\
                           & UHV, UV &  &  & 4.2 & 0.8 \\
\hline
\multirow{2}{*}{SiN$_x$-5} & NH$_3$ & \multirow{2}{*}{4.1} & \multirow{2}{*}{0.8} & 1.7 & 0.7 \\
                           & UHV, UV &    &      & 1.1 & 0.6 \\
\hline
\multirow{2}{*}{SiN$_x$-6} & NH$_3$ & \multirow{2}{*}{1.7} & \multirow{2}{*}{1.0} & 1.1 & 0.9 \\
                           & UHV, UV &     &    & 0.35 & 0.6 \\
\hline
SiO$_x$-1 & UHV, UV & 13.4 & 0.5 & 13.7 & 0.5 \\
\hline
SiO$_x$-2 & UHV, UV & 6.5 & 1.0 & 2.5 & 0.9 \\
\hline
SiO$_x$-3 & UHV, UV & 4.8 & 0.7 & 5.1 & 1.1 \\
\hline
SiO$_x$-4 & UHV, UV & 3.0 & 0.8 & 5.4 & 0.8 \\

\end{tabular}
\caption{\label{tab:tab1}Noise reduction by vacuum and surface treatment. The Table includes results of before/after measurements on six SQUIDs with SiN$_x$ loop encapsulation (SiN$_x$-1...6) and four SQUIDs with SiO$_x$ loop encapsulation (SiO$_x$-1...4). Relative uncertainties in flux noise power spectral density $S_\Phi$(1~Hz) and noise exponent $\alpha$ are 10\% and 25\%, respectively, as determined from repeated measurements following thermal cycling (see Supplement).}
\end{table}

In the case of SQUIDs encapsulated in SiN$_x$, we observe a significant noise reduction both for devices passivated with NH$_3$ and for devices cooled in improved vacuum following UV illumination. Fig. \ref{fig:fig4}a shows before/after spectra from one sample that was baked in the titanium cell and passivated with NH$_3$ using the protocol described above. The flux noise power spectral density at 1 Hz decreases from 8.2~$\mu\Phi_0^2/$Hz to 1.6~$\mu\Phi_0^2/$Hz. In Fig. \ref{fig:fig4}b we show before/after spectra from a device that was subjected to UV illumination and cooled in improved vacuum; here, the flux noise power spectral density at 1 Hz decreases from 1.7~$\mu\Phi_0^2/$Hz to 0.35~$\mu\Phi_0^2/$Hz. We have examined a total of 6 SiN$_x$-encapsulated devices; the results are summarized in the Table. For these devices, we observe a magnetic flux noise level of $3.9 \pm 2.2 \, \mu\Phi_0^2/$Hz at 1 Hz prior to surface treatment, with noise exponent $\alpha = 0.95 \pm 0.17$. Following treatment, we find a noise level $1.7 \pm 1.0 \, \mu\Phi_0^2/$Hz at 1 Hz with noise exponent $\alpha = 0.83 \pm 0.18$. A noise reduction is seen in every SiN$_x$ encapsulated device, with an average reduction in $S_\Phi$(1 Hz) by a factor of 2.8 and a maximum noise reduction by a factor of 5.1. We remark that repeated noise measurements on individual devices (even following thermal cycle to 300~K) show very small variation in the absence of surface modification (see Supplement); the robustness of the noise spectrum to thermal cycling suggests that fixed disorder at the surface dictates how the O$_2$ molecules are adsorbed, or alternatively that strongly bound magnetic species persist to high temperature, providing a noise ``fingerprint'' for each device. The noise suppression observed in the surface treated samples is far beyond what one would expect from typical run-to-run variation. To our knowledge, the $1/f$ flux noise spectral densities measured in our surface-treated nitride devices are the lowest reported in the literature, when the noise is appropriately scaled by the device aspect ratio.

In the case of SiO$_x$-encapsulated devices subjected to UV irradiation under vacuum, no clear noise suppression is seen. We speculate that this is because the UV photon energy of 365 nm (3.4 eV) is large enough to break bonds in the encapsulating oxide, perhaps liberating additional oxygen and providing another path for magnetic contamination.

Our ability to reduce $1/f$ flux noise power by up to a factor of 5 indicates clearly that adsorbates are the dominant source of low-frequency flux noise in our devices. It is reasonable to ask why the noise reduction is not larger. It could be that the remaining noise is still dominated by residual adsorbates, due to the finite partial pressure of O$_2$ in the titanium cell. We measure pressure in the $10^{-9}$~Torr range at the ion pump, and pressure in the cell is likely an order of magnitude higher. Improvements in vacuum could lead to further noise reduction. Once again, the suppression of static spin susceptibility in the Nb SQUID described in Fig. \ref{fig:fig3} is larger than the noise reductions in Al-based devices described in Fig. \ref{fig:fig4} and Table I. This discrepancy suggests to us that the details of the disordered surface of the device play a critical role in dictating the adsorption and/or fluctuation dynamics of the O$_2$ moments. 
Alternatively, it could be that the residual noise is due to some other magnetic states at the buried metal-substrate interface, or at the interface between the metal and its native oxide. 

Our DFT calculations indicate that an O$_2$ molecule adsorbed on Al$_2$O$_3$ (0001) sits atop Al atoms and
has a spin of 1.8 $\mu_B$ that rotates almost freely in the plane perpendicular to the molecular axis (barrier to spin rotation $\sim 10$~mK) \cite{Wang15}. $1/f$ noise results from a distribution of relaxation times \cite{Dutta81} that can arise from spin-spin interactions. DFT finds that neighboring O$_2$ molecules have ferromagnetic exchange, and Monte Carlo simulations show that a distribution of ferromagnetic interactions produces $1/f$ noise consistent with experiment \cite{Wang15}. Surface disorder could change the magnitude and sign of these interactions, affecting the noise exponent $\alpha$; these questions are the focus of ongoing research.



In summary, we find that adsorbed molecular O$_2$ is a dominant source of magnetism in superconducting devices. Suitable surface passivation and improvements in the sample vacuum environment lead to significant reductions in the surface spin susceptibility and low-frequency flux noise power. These results open the door to realization of frequency-tunable superconducting qubits with improved dephasing times.

\begin{acknowledgments}
We thank A. Puglielli and T. Klaus for technical assistance and we acknowledge useful discussions with J. L. DuBois, L. Faoro, L. B. Ioffe, V. Lordi, and J. M. Martinis. This work was supported in part by the U.S. Government under grants W911NF-09-1-0375 and W911NF-10-1-0494 and by a gift from Google, Inc. DFT calculations at UCI (HW and RW) were supported by DOE-BES (Grant No. DE-FG02-05ER46237) and NERSC. Work at Fudan (ZW and RW) was supported by the CNSF (Grant No. 11474056). Work at the Advanced Photon Source, Argonne was supported by the U.S. Department of Energy, Office of Science under Grant No. DEAC02-06CH11357. NIST acknowledges support from the LPS, ARO, and DARPA. 
\end{acknowledgments}

\end{document}


\patchcmd{\section}
  {\centering}
  {\raggedright}
  {}

\title{Supplement to \\  \enquote {Origin and Suppression of $1/f$ Magnetic Flux Noise}}

\author{P. Kumar}
\author{S. Sendelbach}
\altaffiliation[Present address: ]{Northrop Grumman Corporation, Linthicum, Maryland 21203, USA}
\author{M. A. Beck}
\affiliation{Department of Physics, University of Wisconsin-Madison, Madison, Wisconsin 53706, USA}
\author{J. W. Freeland}
\affiliation{Advanced Photon Source, Argonne National Laboratory, Argonne, Illinois 60439, USA}
\author{Zhe Wang}
\affiliation{Department of Physics and Astronomy,
University of California, Irvine, California 92617, USA}
\affiliation{State Key Laboratory of Surface Physics and Key Laboratory
for Computational Physical Sciences, Fudan University, Shanghai 200433, China}
\author{Hui Wang}
\affiliation{Department of Physics and Astronomy,
University of California, Irvine, California 92617, USA}
\affiliation{State Key Laboratory of Surface Physics and Key Laboratory
for Computational Physical Sciences, Fudan University, Shanghai 200433, China}
\author{C. C. Yu}
\author{R. Q. Wu}
\affiliation{Department of Physics and Astronomy,
University of California, Irvine, California 92617, USA}
\author{D. P. Pappas}
\affiliation{National Institute of Standards and Technology, Boulder, Colorado 80305, USA}
\author{R. McDermott}
\email[Electronic address: ]{rfmcdermott@wisc.edu}
\affiliation{Department of Physics, University of Wisconsin-Madison, Madison, Wisconsin 53706, USA}

%
%
%
%
%

\date{\today}
\maketitle
\makeatletter
\renewcommand{\thefigure}{S\@arabic\c@figure}
\renewcommand{\thetable}{S\@Roman\c@table}
\makeatother

This Supplement provides additional information on DFT calculations, Xray data, and flux noise measurements and analysis relevant to the results and conclusions presented in the main text. The document includes the following sections:
\begin{enumerate} [I.]
\item 	 Methodology and parameters of DFT calculations
\item 	 Additional XAS and XMCD data
\item 	 Device fabrication
\item 	 Measurement scheme and experimental setup
\item 	 Dependence of noise on device aspect ratio
\item 	 Dependence of noise on dielectric encapsulation
\item 	 Device aging
\item 	 Run-to-run variation of measurement results
\end{enumerate}

\section{\label{sec:level1}Methodology and parameters of DFT calculations }
\begin{figure}[t!]
\includegraphics[width=5.5in]{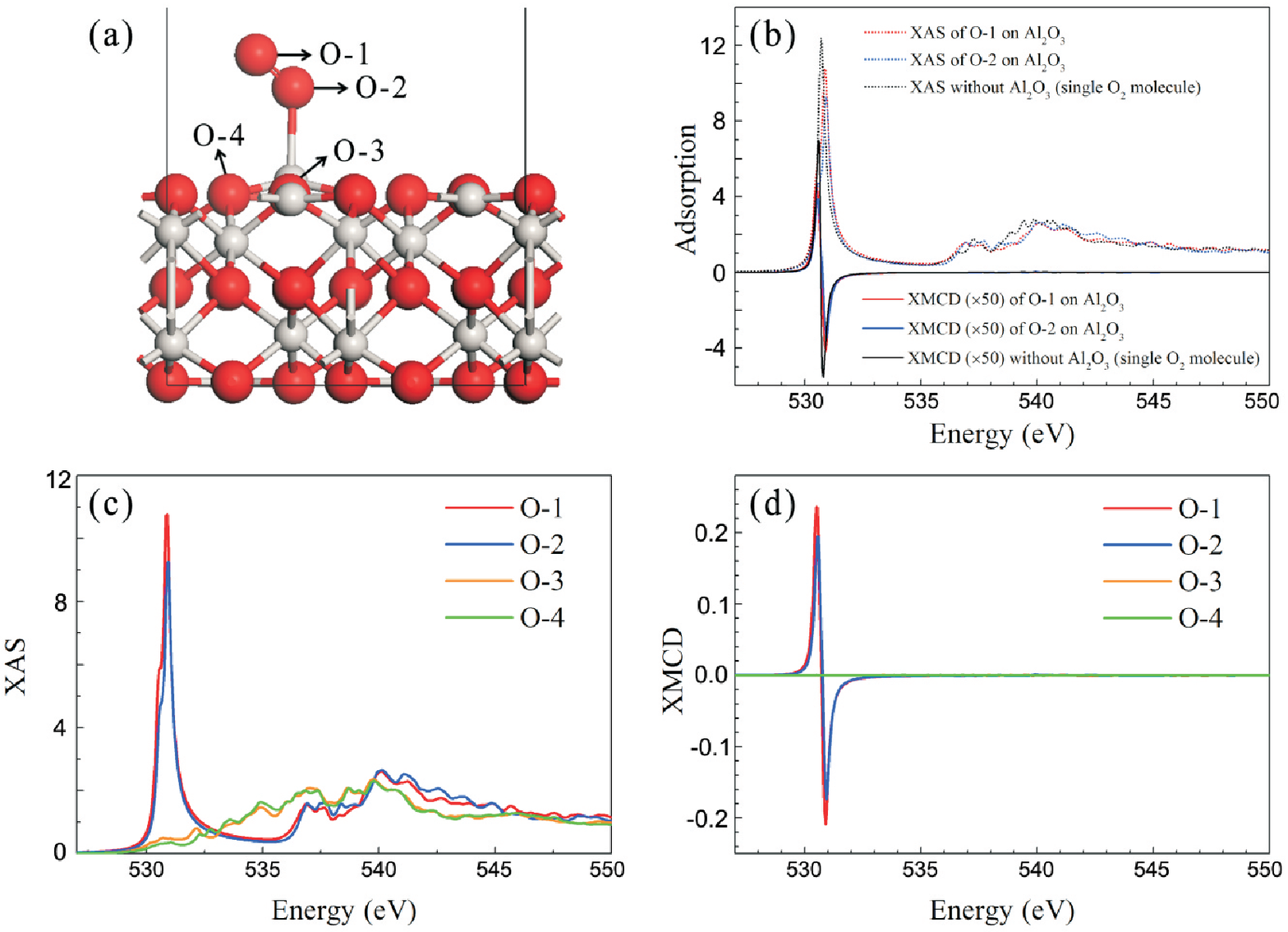}
\vspace*{-0.0in} \caption{(a) Schematic showing an O$_2$ molecule adsorbed on an Al$_2$O$_3$ (0001) surface. (b)-(d) DFT results for O$_2$/Al$_2$O$_3$, with contributions from O$_2$ and lattice oxygen atoms.}
\label{fig:DFT}\end{figure}

Density functional theory (DFT) calculations were carried out using the highly-precise full potential linearized augmented plane-wave (FLAPW) method that has no shape approximation for charge, potential, and wave function expansions \cite{Wimmer81}. We used the generalized gradient approximation (GGA) in the formula of Perdew-Burke-Ernzerhof (PBE) for the description of the exchange-correlation interaction among electrons \cite{Perdew96}. The core electrons were treated fully relativistically, while the spin-orbit coupling term for the valence states was invoked second variationally \cite{Wu99}. Energy cutoffs of 400 Ry and 25 Ry were chosen for the charge/potential and basis expansions in the interstitial region. In the muffin-tin region ($r_{\textrm{Al}}$ = 2.00 a.u. and $r_{\textrm{O}}$ = 1.16 a.u.), spherical harmonics with a maximum angular momentum quantum number of $l_{\textrm{max}}$ = 8 were used for all expansions. Electronic self-consistency was assumed when the root-mean-square differences between the input and output charge and spin densities were less than 1.0~$\times~10{^{-5}}$~$e/(\textrm{a.u.})^{3}$. Numerical convergence of all physical properties against the number of $k$ mesh points based on Monkhorst-Pack in the Brillouin zone (BZ) was carefully examined. Xray magnetic circular dichroism (XMCD) and Xray absorption spectra (XAS) were calculated according to the linear response equations elaborated in reference \cite{Wu99}. All DFT spectra were shifted upward in energy by 20 eV to compensate for the final state effect in the Xray absorption process. DFT results for O$_{2}$ adsorbed on Al$_{2}$O$_{3}$ (0001), with contributions from lattice oxygen atoms, are shown in Fig. \ref{fig:DFT}.

\section{\label{sec:level1}Additional XAS and XMCD data}
\begin{figure}[t!]
\includegraphics[width=5.5in]{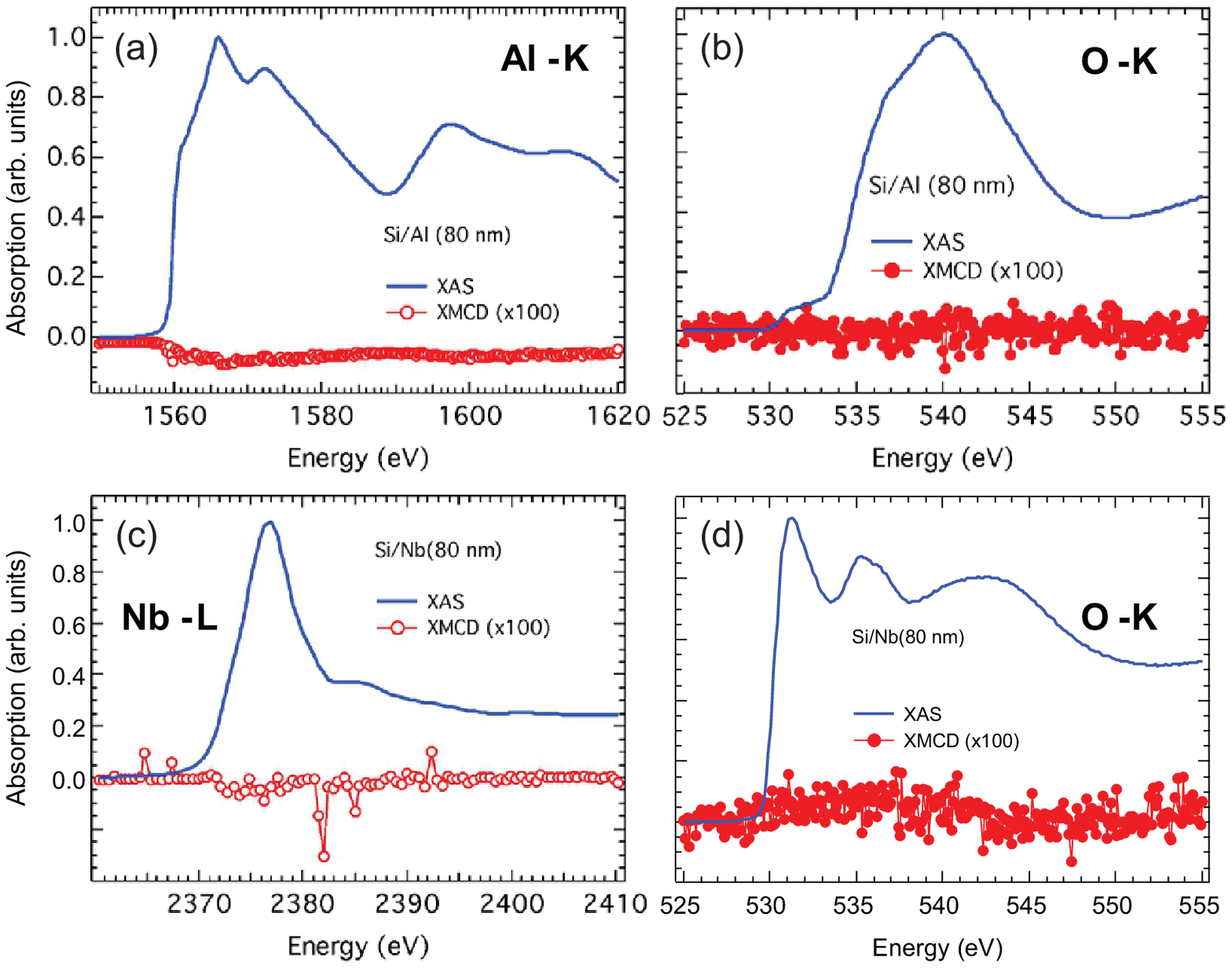}
\vspace*{-0.0in} \caption{XAS and XMCD data for (a) the Al K-edge and (b) the O K-edge in the Al thin film. XAS and XMCD data for (c) the Nb L-edge and (d) the O K-edge in the Nb thin film. No XMCD signal was observed at any of these energies.}
\label{fig:XMCD}\end{figure}

In our initial set of experiments at the APS beamline, we examined the Al and O K-edges in Al films and the Nb L-edge and O K-edge in Nb films and observed no XMCD signal at any of these energies; see Fig. \ref{fig:XMCD}. For these measurements, the Al and the Nb thin film samples were cooled to the cryostat base temperature of 10~K in ultrahigh vacuum (UHV; $P$ < 10$^{-9}$ Torr) conditions.

\section{\label{sec:level1}Device fabrication}
\begin{figure}[t!]
\includegraphics[width=5in]{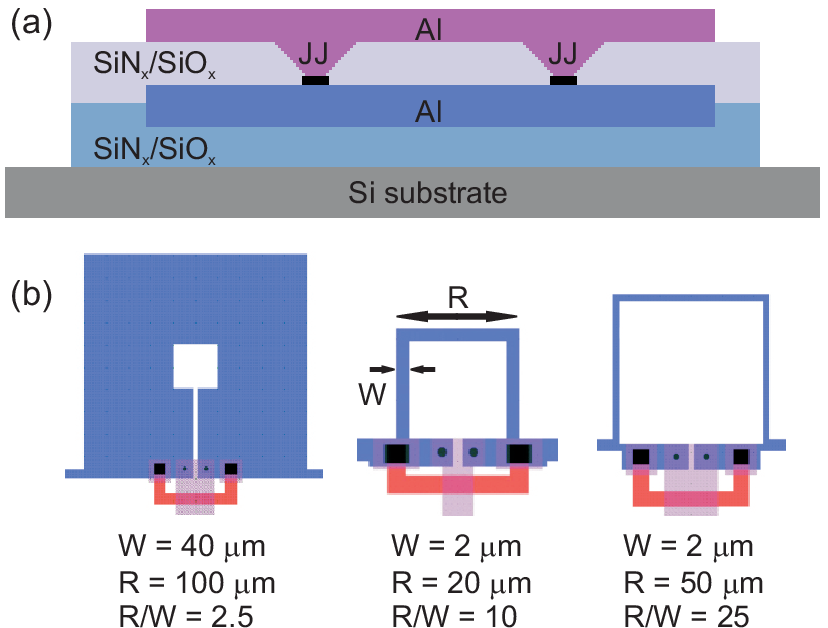}
\vspace*{-0.0in} \caption{(a) Sketch of Al SQUID layer stack. Upper dielectric: PECVD-grown SiN$_{x}$ or SiO$_{x}$. Lower dielectric: thermal SiO$_{x}$ or PECVD-grown SiN$_{x}$. (b) Device layouts showing different aspect ratios $R/W$ used for investigation of dependence of flux noise on geometry.}
\label{fig:device}\end{figure}

The aluminum first-stage SQUIDs were fabricated in the Wisconsin Center for Applied Microelectronics (WCAM) at the University of Wisconsin using a four layer process; a schematic of the device layer stack is shown in Fig. \ref{fig:device}a, and the investigated geometries are shown in Fig. \ref{fig:device}b. The aluminum base layer was deposited on an oxidized or nitrided silicon substrate by sputtering, and the layer was patterned and etched using Transene Etchant A. Next, a silicon nitride (SiN$_{x}$) or silicon oxide (SiO$_{x}$) wiring dielectric layer was grown by plasma enhanced chemical vapor deposition (PECVD). Vias were etched in the wiring dielectric using reactive ion etching (RIE) to define the Josephson junctions. An ion mill was used to remove the native oxide of the base layer, and the aluminum oxide of the tunnel barrier was formed by thermal oxidation in dry O$_{2}$. The Al counter electrode was deposited by sputtering, and the counter electrode traces were defined by a Transene wet etch. Finally, the Pd shunt resistors were formed by electron beam evaporation and liftoff. The junction areas were 2 $\mu$m$^{2}$, and the single-junction critical currents were in the range from 3-5 $\mu$A.

\section{\label{sec:level1}Measurement scheme and experimental setup}
\begin{figure}[t!]
\includegraphics[width=5in]{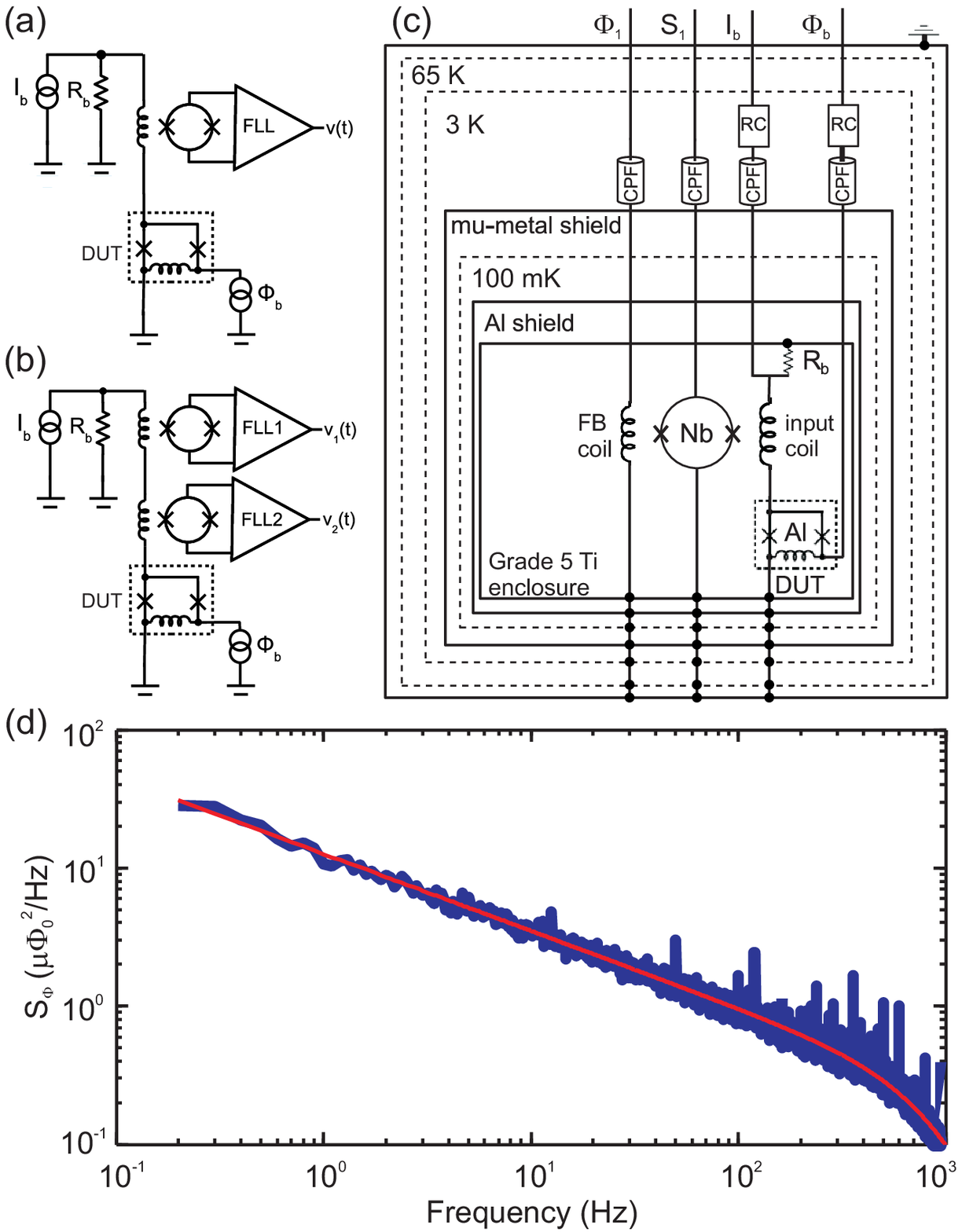}
\vspace*{-0.0in} \caption{Experimental setup and measurement scheme. (a) SQUID amplifier circuit. The Al-based device under test (DUT) is voltage biased and the external flux bias ($\Phi_b$) is tuned to the point of maximum transfer function $dI/d\Phi$; current through the DUT is read out by a second-stage Nb-based SQUID, which is operated in flux locked loop (FLL). (b) Cross-correlation amplifier circuit. (c) Block diagram of measurement wiring in the ADR. The Nb and Al SQUIDs are housed in a grade 5 titanium enclosure and mounted at the cold stage of the ADR. The devices are protected from stray magnetic fields by two layers of superconducting shields (the grade 5 titanium enclosure itself, with $T_c \sim$ 4.5~K, and a second stage of aluminum shielding, with $T_c \sim$ 1.2~K) and by a single layer of cryogenic mu-metal. All bias lines are filtered with copper powder filters (CPF) at the 3~K stage. Additional $RC$ filters are used for the bias lines ($I_b$ and $\Phi_b$)  of the DUT at the 3~K stage. (d) Example noise spectrum of an SiO$_{x}$-encapsulated Al SQUID showing fit to the spectrum (red line).}
\label{fig:measure}\end{figure}

\textit{Noise measurements} - Flux noise measurements were performed on a series of Al-based SQUID devices. Devices were cooled to 100~mK in an adiabatic demagnetization refrigerator (ADR). The device under test was biased with a voltage, and fluctuating current through the device was measured by a second stage Nb-based SQUID incorporating an integrated multiturn input coil. In some cases, two second stage SQUIDs were used to simultaneously monitor the noise of the device under test, and the cross spectrum of the two measurement SQUIDs was used to suppress the added noise of the second stage. The measurement configurations are shown in Fig. \ref{fig:measure}a-b, and a block diagram of the experimental wiring is shown in Fig. \ref{fig:measure}c. Multiple layers of magnetic shielding were used to isolate the SQUIDs from ambient field fluctuations. The devices were protected from high-frequency interference by heavy filtering on the bias lines to the SQUIDs. All wires to the experimental stage incorporated copper powder filters (CPFs) at 3~K. Additional inline $RC$ filters were employed on the voltage- and flux-bias lines of the Al first-stage SQUID. Each second stage SQUID was operated in a flux-locked loop (FLL); the FLL output was passed through an anti-aliasing filter and digitized for further analysis. The flux noise power spectral density of the device under test was computed for each time series, and hundreds of such spectra were averaged together. We fit the averaged flux noise spectra to the functional form $S_{\Phi}$($f$)~=~$A/f^{\alpha}$ + $B$ to extract the noise power $A$ and exponent $\alpha$. A typical noise spectrum for an SiO$_{x}$-encapsulated device is shown in Fig. \ref{fig:measure}d along with the fit.\\

\textit{Grade 5 titanium enclosure} - The noise measurements were conducted in a specially designed hermetic sample enclosure made from grade 5 titanium alloy (Ti-6Al-4V); see Fig. 2 in the main text. This alloy has excellent UHV properties due to its low outgassing and its hardness, allowing realization of all-metal conflat seals. Moreover, the material is compatible with high-bandwidth weld-in hermetic SMA connectors, enabling both sensitive low-frequency measurements of flux noise and microwave-frequency control and readout needed for qubit experiments. Finally, grade 5 titanium superconducts at around 4.5 K, providing a magnetic shield for sensitive thin-film superconducting devices.\\

\textit{Experimental protocol for study of surface treatment} - Devices were first mounted in the titanium enclosure and baseline data was collected: here, devices were not subjected to any surface modification and the titanium enclosure was not evacuated prior to installation in the ADR. In a subsequent cooldown, the same device was remeasured following improvement of the sample vacuum environment and/or surface modification by NH$_{3}$ passivation or UV illumination. The vacuum treatment included bakeout at 120$\degree$C under vacuum for several days to achieve pressure at the ion pump around 10$^{-9}$ Torr. The UV-treated samples were irradiated under vacuum with a UV LED (300 mW at 365 nm); the UV LED was mounted on the conflat gasket a few millimeters from the sample using UHV-compatible epoxy. In some cases, a nonevaporable getter (NEG) pill (SAES Inc.) was activated in a separate chamber and transferred into the sample enclosure under vacuum prior to pinchoff of the vacuum cell. The NEG pill provides continuous pumping in the sample cell following pinchoff. In the case of NH$_{3}$ passivation, the titanium enclosure was first baked and evacuated until UHV conditions were reached; the cell was subsequently backfilled with 100 Torr of NH$_{3}$ before pinchoff. Flux noise power spectra of the surface treated devices were measured and compared to the baseline data.\\

\section{\label{sec:level1}Dependence of noise on device aspect ratio}
\begin{figure}[t!]
\includegraphics[width=5in]{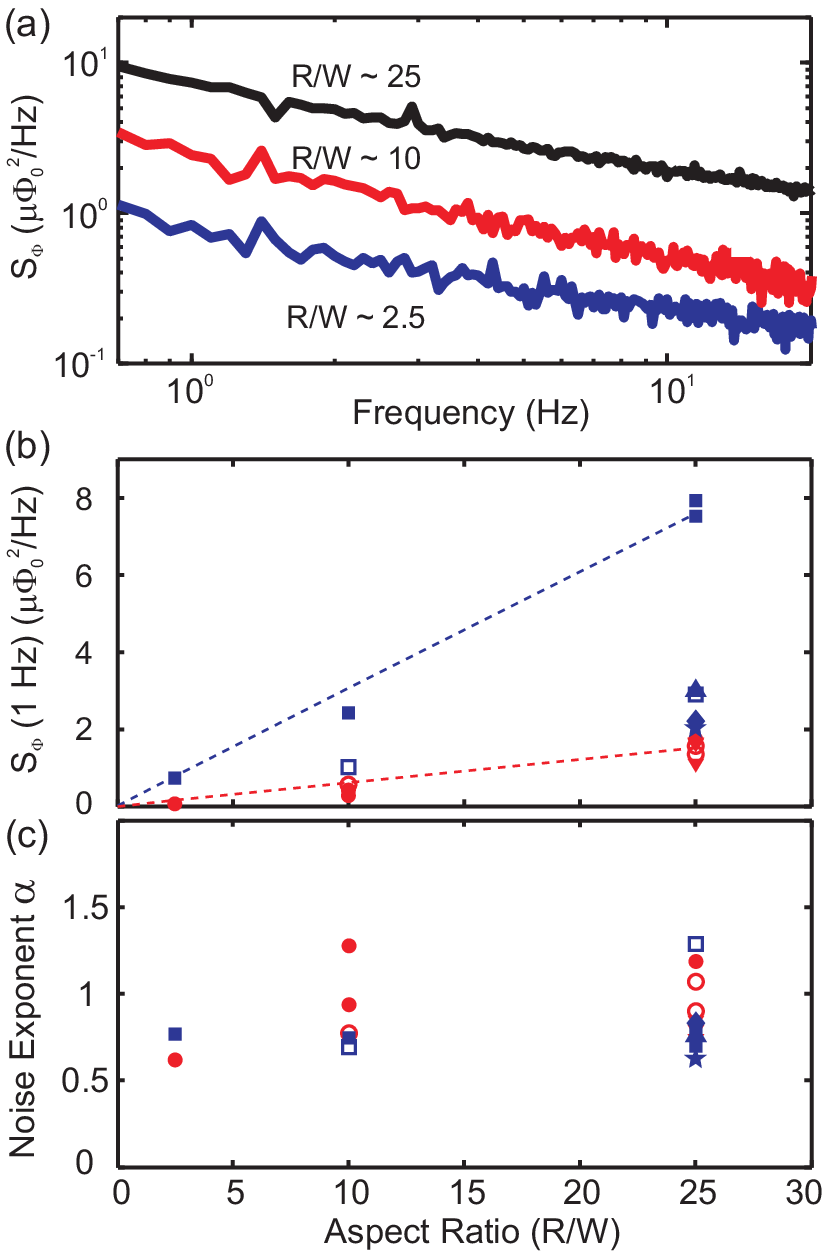}
\vspace*{-0.0in} \caption{Dependence of noise power spectral density on device aspect ratio and dielectric encapsulation. (a) Flux noise power spectra of three SiO$_x$-encapsulated devices with aspect ratio $R/W$= 2.5, 10, and 25; the linear scaling of noise power with device aspect ratio is clearly seen. (b) Dependence of flux noise power spectral density $S_\Phi$(1 Hz) on aspect ratio. The data span 8 different wafers incorporating both SiO$_{x}$ (blue points) and SiN$_{x}$ (red points) dielectric encapsulation. For devices cofabricated on a single wafer (indicated by identical symbols), the linear scaling of flux noise with aspect ratio is clearly seen (blue and red dashed lines). (c) Flux noise exponent $\alpha$ versus aspect ratio; no clear dependence is seen.}
\label{fig:AR}\end{figure}

We observe a clear dependence of SQUID flux noise power spectral density on the aspect ratio of the SQUID loop, i.e., the ratio of loop width $R$ to trace width $W$. This can be understood simply from reciprocity: the coupling of individual spins to the SQUID loop is proportional to the surface magnetic field at the spin location due to a test current that is injected in the SQUID loop \cite{Bialczak07,Anton13,LaForest15}. The investigated geometries are shown Fig. \ref{fig:device}b. We expect that flux noise power spectral density should scale linearly with $R/W$, apart from log corrections. In Fig. \ref{fig:AR}a we show flux noise power spectra of three SiO$_{x}$-encapsulated devices cofabricated on a single wafer; the expected scaling of noise power with aspect ratio is clearly seen. In Fig. \ref{fig:AR}b, we plot the flux noise power spectral density versus aspect ratio for a range of devices fabricated on 8 different wafers and incorporating both SiO$_{x}$ and SiN$_{x}$ encapsulation. For devices cofabricated on a single wafer, the linear scaling of flux noise with aspect ratio is clearly seen (red and blue dashed lines). In Fig. \ref{fig:AR}c, we show the dependence of flux noise exponent $\alpha$ on aspect ratio. No clear dependence is seen.

\section{\label{sec:level1}Dependence of noise on dielectric encapsulation}
Our measurements of the $1/f$ magnetic flux noise in these SQUIDs show a clear dependence of the noise power on the materials used to encapsulate the device loop, even in the absence of surface treatment. The data in Fig. \ref{fig:AR}b-c are from devices encapsulated in SiO$_{x}$ (blue points) and in SiN$_{x}$ (red points). We see significantly lower noise in the SiN$_{x}$-encapsulated devices, although the wafer-to-wafer variation in the noise from SiO$_{x}$-encapsulated devices is large, and some oxide-encapsulated devices show noise comparable to nitride-encapsulated devices. Preliminary results of our DFT calculations indicate there may be a high energetic barrier to reorientation of the O$_2$ magnetic moment when the molecule is adsorbed on a nitride-based surface as opposed to an oxide-based surface. Thus, adsorbed O$_2$ remains magnetically active down to millikelvin temperatures on an oxide surface, whereas the magnetic moment of O$_2$ starts to freeze out at higher temperatures on a nitride surface, leading to a reduction in noise.

\section{\label{sec:level1}Device aging}
The devices described in this manuscript were fabricated in 2012-2013. The study of the impact of device aspect ratio on noise was performed during that same interval; however, investigation of the effect of surface treatment was performed much later, in the time frame 2015-2016. Devices were not stored in a dry box or in a controlled environment. In several cases, devices that were characterized soon after fabrication were remeasured several years later. We find that the noise of the nitride-encapsulated devices increased significantly over the course of this multi-year study, suggesting a slow aging of the disordered device surface or a long timescale for the adsorption of strongly-bound magnetic species at 300~K. Results are shown in Table SI.\\

\restylefloat{table}
\begin{table}[!t]
\begin{center}
\floatstyle{plaintop}
\caption{Flux noise power spectral density $S_\Phi$(1 Hz) and noise exponent $\alpha$ for several SiN$_{x}$- and SiO$_{x}$-encapsulated devices that were measured shortly after fabrication and subsequently remeasured years later (in all cases, device aspect ratio $R/W$ = 25). The noise in the SiN$_{x}$-encapsulated devices increased significantly over time.}
\begin{tabular}{ |c|c c|c c| }
 \hline
\multicolumn{1}{|c|}{} & \multicolumn{4}{|c|}{Measurement Period}\\
\multicolumn{1}{|c|}{} &  \multicolumn{2}{|c|}{2012-2013} & \multicolumn{2}{|c|}{2015-2016}\\
 \hline
Device   & S$_\Phi$ (1 Hz)  &  $\alpha$\, & S$_\Phi$ (1 Hz)  &  $\alpha$\, \\
encapsulation   & ($\mu\Phi_{0}^2$/Hz) &  &  ($\mu\Phi_{0}^2$/Hz) & \\
 \hline
 \multirow{6}{*}{SiN$_x$}  & 1.1 &   0.7 \,& 1.7 &   1.0\,\\
                      & 1.4 &   1.1 \,& 2.0 &   1.0\,\\
                      & 1.6 &   0.9 \,& 2.8 &   1.0\,\\
                      & 1.6 &   0.8 \,& 4.1 &   0.8\,\\
                      & 1.7 &   0.8 \,& 4.4 &   0.7\,\\
                      & 1.7 &   1.2 \,& 8.2 &   1.2\,\\
 \hline
 \multirow{4}{*}{SiO$_{x}$}   & 2.9 &   1.3 \,& 3.0 &   0.8\,\\
                      & 3.0 &   0.8 \,& 4.8 &   0.7\,\\
                      & 7.5 &   0.7 \,& 6.5 &   1.0\,\\
                      & 7.9 &   0.8 \,& 13.4 &   0.5\,\\
 \hline
\end{tabular}
\end{center}
\end{table}

\section{\label{sec:level1}Run-to-run variation of measurement results}
\begin{figure}[!t]
\includegraphics[width=7in]{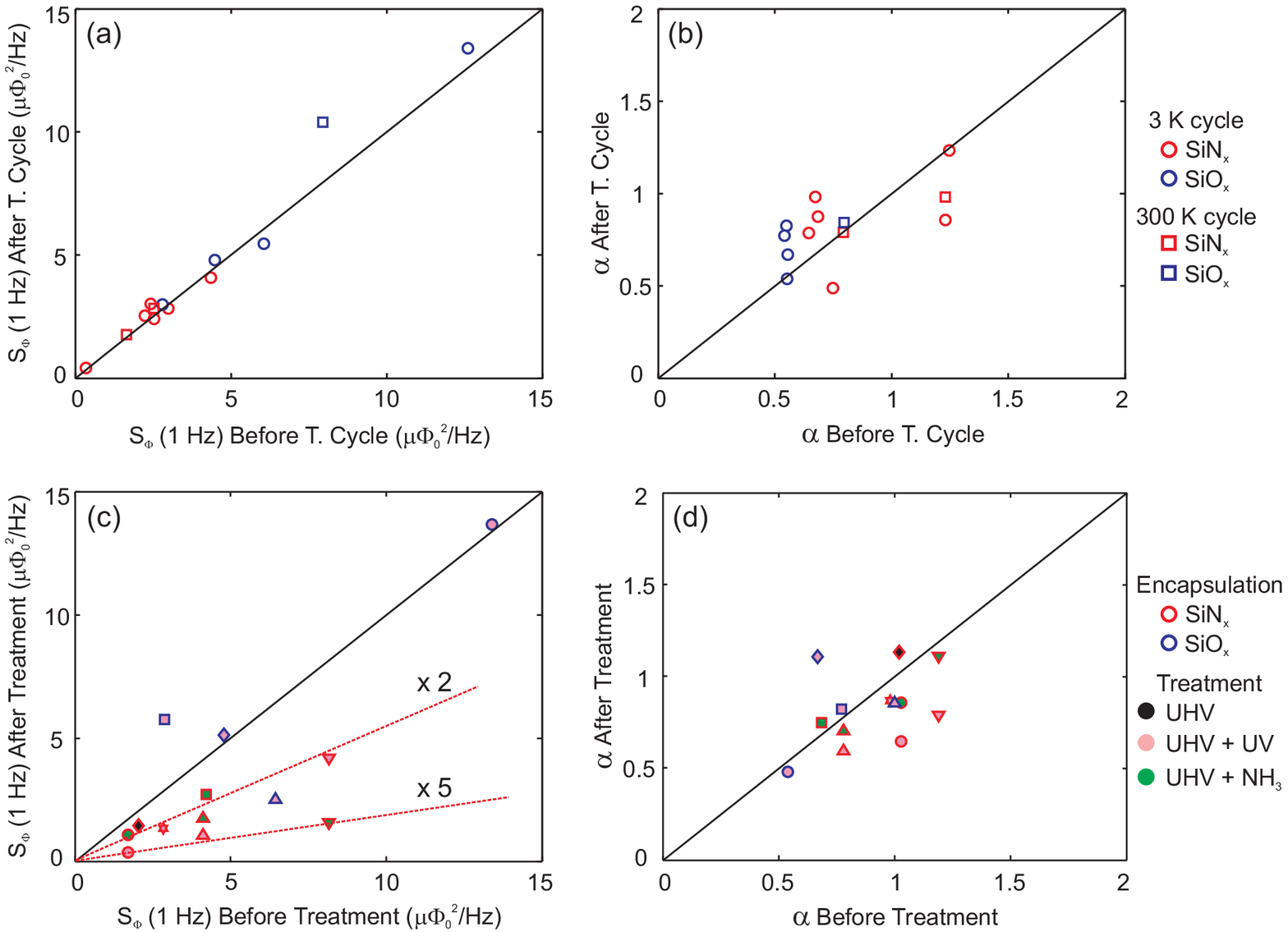}
\vspace*{-0.0in} \caption{(a) Flux noise power spectral density $S_\Phi$(1 Hz) and (b) noise exponent $\alpha$ before and after thermal cycle to 3~K and 300~K, without surface treatment. Blue points correspond to SiO$_x$-encapsulated devices, and red points correspond to SiN$_x$-encapsulated devices. (c)-(d) Analogous before/after plots of $S_\Phi$(1 Hz) and $\alpha$ for devices subjected to various surface treatments. Here, the blue border corresponds to SiO$_x$ encapsulation, the red border corresponds to SiN$_x$ encapsulation, and the fill color identifies the surface treatment (black: UHV alone; pink: UHV+UV; green: UHV+NH$_{3}$). An average reduction of $\sim$ 2.8 in noise power spectral density is seen, with maximum noise suppression of 5.1. The red dashed lines in (c) correspond to noise suppression by a factor of 2 and 5.}
\label{fig:Tcycle}\end{figure}

In a typical measurement cycle, devices were characterized multiple times in an ADR at 100~mK, with repeated thermal cycles to 3~K between measurement runs in order to re-charge the ADR magnet. We observe 10\%  scatter in the flux noise power spectral density $A$ and 29\% scatter in the noise exponent $\alpha$ from run to run following thermal cycle to 3~K. In three cases, we re-measured devices following thermal cycle to 300~K and exposure of the device to atmosphere prior to the second cooldown. For these cases, we observe 7\% variation in the noise power spectral density and 13\% variation in the noise exponent. In Fig. \ref{fig:Tcycle}a-b, we present data on the run-to-run variation in device noise. The robustness of the flux noise power spectral density suggests to us that fixed disorder at the surface dictates the adsorption and fluctuation dynamics of the O$_{2}$ spins. For comparison, in Fig. \ref{fig:Tcycle}c-d we show the same before/after data for devices subjected to surface treatment. As discussed in the main text, we observe a significant noise suppression in nitride-encapsulated devices subjected to the various surface treatments, far beyond the typical run-to-run variation of the measurement. The average noise suppression observed in surface-treated nitride samples is $\sim$ 2.8 times, with a maximum noise suppression factor of $\sim$ 5.1. (All data shown in this figure are from high aspect ratio devices, $R/W=25$).
\newpage

